# Text Data Mining from the Author's Perspective:
## Whose Text, Whose Mining, and to Whose Benefit?

Christine L. Borgman
Distinguished Professor & Presidential Chair in Information Studies
Director, Center for Knowledge Infrastructures
University of California, Los Angeles
Christine.Borgman@ucla.edu
http://www.christineborgman.info

Forum Statement: Data Mining with Limited Access Text: National Forum. (2018).
https://publish.illinois.edu/limitedaccess-tdm/

Researchers have sought technical access to proprietary databases of published materials since the earliest days of online databases in the latter 1970s, yet publishers continue to write university contracts based only on human readership. By the time of Google Books and the associated author lawsuits, ca 2005, we learned that publishers wished to restrict "non-consumptive use" of scholarly content (Duguid, 2007; Leetaru, 2008; Nunberg, 2011). Throughout this period, the move toward open access to journal articles accelerated, with arXiv launching in 1991 (Ginsparg, 2011) and PubMed Central in 2000 ("PMC Overview," 2018). Numerous other discipline-specific preprint servers, institutional repositories, and commercial services designed to distribute or redistribute open access versions of scholarly publications have been launched since. Concurrently, open access to publications became mandatory or highly recommended by many funding agencies and universities, in the U.S. and abroad.

Text data mining has become an essential part of scholarship in a world of "big data" where integrating disparate content can lead to new insights. More than 30 years ago, Don Swanson (1986) demonstrated the power of data integration to reveal "undiscovered public knowledge." Swanson's labor-intensive manual methods have been supplanted by machine searching, yet user interfaces and publisher contracts continue to be predicated on assumptions of human-computer interaction. As *The Hague Declaration on Knowledge Discovery in the Digital Age* states, "the right to read is the right to mine" (LIBER, 2015). The *Public Library of Science*, acting on this principle of the Hague Declaration, provides mining access to its corpus of more than 200,000 articles from all of its journals (Public Library of Science, 2018). The FAIR principles (Findable, Accessible, Interoperable, Reusable) for data sharing are similarly based on machine access to scholarly content (Wilkinson et al., 2016).

Given the many technical, social, and policy shifts in access to scholarly content since the early days of text data mining, it is time to expand the conversation about text data mining from concerns of the researcher wishing to mine data to include concerns of researcher-authors about how their data are mined, by whom, for what purposes, and to whose benefits.

This snippet from a recent Data Science Newsletter reflects some of the concerns expressed by authors (Noren & Stenger, 2018):

> **Elsevier** may be shifting its business model to become a data provider to scientists which is different than being a journal publisher. Elsevier execs **were at Harvard's Data Science Initiative**, where they are building intellectual partnerships to explore how this new model could work. **Critics** are warning scientists to heed the internet-age adage: if you aren't the customer, you are the product. With Elsevier (and **Facebook**), the user will be both the customer and the product, to the tune of a 30% profit margin arguably due to the fact that the writing, reviewing, and much of the editorial work are done by volunteers. As we noted last week, **Germany is having none of that anymore**, demanding that all of the articles published by scientists funded by the German state be made freely available to anyone in the world. That's an ongoing battle (Bert, 2017; Schiermeier, 2018; Smith, 2018).

Proponents of open access publishing tend to express one of two goals: (1) democratizing access to knowledge or (2) limiting the role of big publishers in controlling access to scholarly content, and in their ability to charge high fees to universities and readers (Borgman, 2007; Harnad, 1991, 1999, 2005; Suber, 2012; Willinsky, 2006). These competing views are conflated in recent developments, with proprietary publishers charging several thousand dollars (or euros) to make a single article open access, and with large publishers such as Elsevier purchasing independent, community-based open access venues such as SSRN and Bepress. Open access is not turning out to be the information commons that was envisioned by its pioneers (Benkler, 2004; Hess & Ostrom, 2007; Kranich, 2004; Lessig, 2001; O'Sullivan, 2008; Reichman, Dedeurwaerdere, & Uhlir, 2009; Reichman, Uhlir, & Dedeurwaerdere, 2016).

The rise of bibliometrics and "altmetrics" to assess scholarly productivity in the academy also has authors worried about who can mine their content and for what uses. Despite a long history of scholarship that demonstrates the meaninglessness of most of these numbers in assessing actual impact or long-term value of an individual's contributions, the uses of these metrics continue to proliferate (Borgman, 1990, 2016; Cronin, 2005; Cronin & Atkins, 2000; Cronin & Sugimoto, 2014, 2015). Calculations of citations, H-index, and other indicators vary widely between common sources such as Web of Science, Google Scholar, and Scopus, due to differences in editorial coverage, algorithms, and methods used by those attempting to mine these databases.

Whether mining a corpus for bibliometrics, textual content, images, or numeric data, the bibliographic descriptions are essential metadata. Original articles typically provide accurate bibliographic descriptions, and may also include "please cite as" instructions. However, references to published articles, which are essential for bibliometrics or for integrating content across databases, are inherently dirty data due to the vagaries of how authors create reference lists. A bibliography in a journal article is far from the "necessary and sufficient" set of citations that might be assumed by bibliometric evaluations. Rather, it is often an idiosyncratic list of familiar sources, compiled based on what is handy when the publication is submitted. Too few authors are bibliographic purists who verify middle initials, dates, DOIs, and page, volume, and issue numbers (Borgman, 2015, 2016). Complicating matters further is the lack of agreement on bibliographic styles. At last count, Zotero offered about 9,000 journal styles for referencing, representing about 1750 unique bibliographic styles ("Zotero Style Repository," 2018).

As universities automate academic personnel processes, faculty dossiers also become rich sources of bibliographic data. These records tend to be more accurate than citation lists because authors have a vested interest in establishing claims to their oeuvre. Faculty are becoming concerned about who has access to these data in machine-readable form, and how the data can be mined for making decisions about their careers, their departments, and their fields. Privacy issues abound. Individuals can give informed consent for specific uses of specific data, such as a dossier for a personnel action. When universities employ these data for other purposes or share them with external parties such as the publishers who own some of these dossier systems, privacy and academic freedom issues also arise. Following an extensive analysis of data governance issues at UCLA, for example, operational structures were developed to address the following concerns ("Data Governance Task Force: Final report and recommendations," 2016):

- When data are used to make decisions about people
- When data are collected about people without their knowledge or consent
- When data about people are used in unexpected ways without subjects' knowledge or consent (e.g., new applications of data or systems; mining, analysis, and aggregation)
- When data are used for evaluation purposes
- When data are shared with external entities, whether with research partners or through service contracts with the private sector

A related concern is the ability of publishers to surveil uses of scholarly materials. Ownership of intellectual property carries a large set of rights and responsibilities, some which are associated with privacy protection and intrusion. Corporate owners of scholarly publishing, mass media, and social media content deploy "digital rights management" (DRM) technologies to track uses and users in minute detail. These technologies have eroded traditional protections of privacy and intellectual freedom in libraries and other domains (Cohen, 1996; Lynch, 2017). Universities have special responsibilities for managing their intellectual property in ways that protect the privacy of their communities and minimize harm (Borgman, 2018). The Hague Declaration also includes a principle that "providers of content should respect the intellectual privacy of individual readers" (LIBER, 2015).

Scholarly publishers are becoming data services vendors, entering new markets by acquiring companies in multiple sectors of the information economy. In doing so, they follow the successful business models of Alphabet/Google, Facebook, and Amazon in aggregating vast amounts of data about people's lives. To the consumer, they promote the advantages of improving user experience with intelligent adaptation. To their business clients and investors, they promote the advantages of predictive analytics that can be deployed to strategic advantage. In the academic community, predictive analytics are being used to assess the performance of students and faculty, departments, universities, journals, research programs, and much more. The concentration of data by a few large players gives them a "god's eye view" of their domains, with minimal oversight or regulation ("The world's most valuable resource is no longer oil, but data," 2017). Only gradually are scholarly authors coming to realize that if you are not at the table, you are on the menu. Authors must have a seat at the table in this national (and international) forum on *Data Mining with Limited Access Text.* Privacy, autonomy, and academic freedom are at stake. Let the first question be "Whose Text, Whose Mining, and to Whose Benefit?"

**Acknowledgement:**

Thanks to Michael Scroggins, UCLA, for comments and discussion on an earlier draft.